\documentclass[%
reprint,
amsmath,amssymb,
aip,
]{revtex4-2}

\usepackage{dcolumn}
\usepackage{bm}
\usepackage{hyperref}
\hypersetup{
    unicode=false,  
    pdftoolbar=true,
    pdfmenubar=true,
    pdffitwindow=false,
    pdfstartview={FitH},
    pdfauthor={Eero Hirvijoki},
    pdfcreator={Eero Hirvijoki},
    pdfnewwindow=true,
    colorlinks=true,
    linkcolor=blue,
    citecolor=blue,
    filecolor=blue,
    urlcolor=blue,
    breaklinks=true
}

\begin{document}

\title{Metriplectic foundations of gyrokinetic Vlasov--Maxwell--Landau theory}

\author{Eero Hirvijoki}
\email{eero.hirvijoki@aalto.fi}
\affiliation{Department of Mechanical Engineering, Aalto University, Finland}

\author{Joshua W. Burby}
\affiliation{Los Alamos National Laboratory, Los Alamos, New Mexico 87547, USA}

\author{Alain J. Brizard}
\affiliation{Department of Physics, Saint Michael's College, Vermont 05439, USA}

\date{\today}
\begin{abstract}
This letter reports on a metriplectic formulation of collisional, nonlinear full-$f$ electromagnetic gyrokinetic theory compliant with energy conservation and monotonic entropy production. In an axisymmetric background magnetic field, the toroidal angular momentum is also conserved. Notably, a new collisional current, contributing to the  gyrokinetic Maxwell--Amp\`ere equation and the gyrokinetic charge conservation law, is discovered.
\end{abstract}

\maketitle

\newcommand{\ehat}{\bm{\hat{e}}}
\newcommand{\bhat}{\bm{\hat{b}}}
\newcommand{\rhohat}{\boldsymbol{\hat{\rho}}}
\newcommand{\perphat}{\boldsymbol{\hat{\perp}}}
\newcommand{\lie}{\mathcal{L}}

The theoretical foundations of plasma physics are based on two sets of complementary formulations that are either kinetic or fluid and can represent either collisionless or collisional (dissipative) systems. When the formulations are collisionless, the associated Lagrangian and Hamiltonian structures (see a review by Morrison \cite{Morrison_2017}) play a role in extracting conservation laws and guide, e.g., the development of modern numerical simulation methods \cite{Squire-Qin-Tang-PIC:2012PhPl,Evstatiev-shadwick:2013JCoPh,Shadwick-Stamm-Evstatiev:2014PhPl,Stamm-Shadwick-Evstatiev:2014ITPS,Xiao-et-al-kinetic:2015PhPl,He-et-al-Hamiltonian-splitting:2015PhPl,Qin-et-al:2016NucFu,Xiao-et-al-fluid:2016PhPl,Kraus-et-al:2017JPlPh,Xiao-et-al:2018PlST,Xiao-Qin-6d-tokamak:2020arXiv,Hirvijoki_Kormann_Zonta:2020PoP}. When the formulations include collisional effects, the properties of the collision operator ought to guarantee that the irreversible plasma evolution satisfies the laws of thermodynamics~\cite{Helander_Sigmar}.

The modern theory of gyrokinetics \cite{Brizard_Hahm:RMP}, that is used in the challenging task of investigating the turbulent dynamics of a magnetically-confined plasma in a realistic geometry, has a solid foundation in the collisionless regime. The Lagrangian (variational) structure of the theory \cite{Sugama_gk_field_theory:2000,Brizard:2000} enables deriving conservation laws \cite{Hirvijoki_2020,Brizard:2021} that provide useful verification tests for numerical algorithms. The accompanying, rather recently developed Hamiltonian structure \cite{Burby_Brizard_Morrison_Qin:2015PhLA,Brizard_bracket_2021}, on the other hand, establishes a transparent formulation of the full-$f$ electromagnetic gyrokinetics in terms of genuine dynamical variables that are the gyrocenter phase-space density distributions $F_s$ of species $s$, the electromagnetic displacement field $\bm{D}$, and the perturbation magnetic field $\bm{B}_1$. The existence of these formalisms owes to the dynamical reduction of the original Vlasov--Maxwell theory with the Lie-transform perturbation method \cite{hori1966theory,Deprit:CM1969,Dewar:1976,Littlejohn:JMP1979,CARY1981129,Littlejohn:PoF1981,Littlejohn:JMP1982,Brizard_2006} at the level of the action integral.

Currently, no similar systematic treatment exists for a gyrokinetic version of the nonlinear Landau collision operator compatible with the gyrokinetic Vlasov--Maxwell system. Despite several attempts at constructing collision operators for gyrokinetic applications \cite{Brizard_2004PhPl,Li_Ernst:PRL2011,Madsen:PRE2013,Hirvijoki_Brizard_Pfefferle:2017JPlPh,Pan_Ernst:PRE2019,frei_ball_hoffmann_jorge_ricci_stenger_2021}, no full-$f$ gyrokinetic Vlasov--Maxwell--Landau field theory has yet been presented that would conserve energy, produce entropy monotonically, and conserve the toroidal angular momentum functional in an axially symmetric background magnetic field $\bm{B}_0$. Only the full-$f$ electrostatic theory has been properly covered \cite{Burby_Brizard_Qin_collisions:2015PhPl,Hirvijoki_Burby_collisions:2020PhPl}, and the succesfull $\delta f$-formulations \cite{Abel_et_al_theory:POP2008,Sugama_et_al:PoP2009,Abel_et_al:2013RPPh,Sugama_et_al:PoP2019} rely on various model linearized collision operators. Exploiting the general framework of metriplectic dynamics \cite{Kaufman:1982fl,Kaufman:1984fb,Morrison:1984ca,Morrison:1984wu,Grmela:1984dn,Grmela:1984ea,Grmela:1985jd,Morrison:1986vw}, this letter proposes a theory for the electromagnetic full-$f$ case, securing the as-of-yet-missing closed form expression for the gyrokinetic Landau collision operator. A notable outcome of the new theory is the appearance of a collisional current in the Maxwell--Amp\`ere equation and the gyrokinetic charge conservation law. This new current is shown to be mandatory for the Gauss's law for $\bm{D}$ and the charge conservation to remain valid. It arises due to the non-zero spatial components of collisional flux in gyrocenter coordinates \cite{Brizard_2004PhPl}. Next, the details, leading to these realizations, will be presented.

Following \cite{Burby_Brizard_Morrison_Qin:2015PhLA}, we assume (without loss of generality) gyrocenter coordinates have been found such that the single-gyrocenter Hamiltonian is given by $H_s = K_s[\bm{E}_1,\bm{B}_1] + q_s\,\Phi$, where $\Phi = \Phi(\bm{X})$ denotes the electrostatic potential evaluated at the gyrocenter position, and the gyrocenter kinetic energy $K_s[\bm{E}_1,\bm{B}_1]$ is a functional of the electromagnetic field. (We refer readers to \cite{Burby_Brizard_2019} for details on how explicit dependence of $K_s$ on the electromagnetic potentials can be avoided.) We then use the gyrocenter kinetic energy $\mathcal{K}[\bm{E}_1,\bm{B}_1] = \sum_s \int K_s[\bm{E}_1,\bm{B}_1]\,F_s$ to define the gyrokinetic constitutive law,
\begin{align}
    \bm{D} = \bm{E}_1 - 4\pi\frac{\delta\mathcal{K}}{\delta \bm{E}_1},\label{GK_constitutive}
\end{align}
which relates the electric field $\bm{E}_1$ and the displacement field $\bm{D}$ in a manner first described in \cite{Morrison_2013}. Going forward, we will always assume $\bm{E}_1$ is a definite functional of $\bm{D}$, $\bm{B}_1$, and $F_s$, implicitly defined by \eqref{GK_constitutive}. While no specific expression for the gyrocenter kinetic energy function $K$ is chosen here, it is noted that an explicit expression for $\bm{E}_1[F,\bm{D},\bm{B}_1]$ is available at the drift-kinetic limit (see, e.g., \cite{Zonta_et_al_dispersion_relation:PoP2021}). The Hamiltonian functional for the gyrokinetic Vlasov--Maxwell system is
\begin{align}\label{eq:hamiltonian_functional}
    \mathcal{H}[F,\bm{D},\bm{B}_1]&=\sum_s\int_Z F_s K_s[\bm{E}_1,\bm{B}_1]
    + \frac{1}{4\pi}\int_{X}\bm{E}_1\cdot\bm{D}
    \nonumber
    \\
    &
    -\frac{1}{8\pi}\int_{X}\left(|\bm{E}_1|^2-|\bm{B}_0+\bm{B}_1|^2\right),
\end{align}
and the associated functional Poisson bracket is \cite{Burby_Brizard_Morrison_Qin:2015PhLA,Brizard_bracket_2021}
\begin{align}\label{eq:functional_poisson_bracket}
    &[\mathcal{A},\mathcal{B}]=\sum_s\int_Z F_s\left\{\frac{\delta \mathcal{A}}{\delta F_s},\frac{\delta \mathcal{B}}{\delta F_{s}}\right\}_s
    \nonumber
    \\
    &
    +\sum_s4\pi q_s\int_Z F_s\left(\frac{\delta\mathcal{B}}{\delta \bm{D}}\cdot\left\{\bm{X},\frac{\delta \mathcal{A}}{\delta F_s}\right\}_s-\frac{\delta\mathcal{A}}{\delta \bm{D}}\cdot\left\{\bm{X},\frac{\delta \mathcal{B}}{\delta F_s}\right\}_s\right)\nonumber
    \\
    &+\sum_s 16\pi^2q_s^2\int_Z F_s\frac{\delta \mathcal{A}}{\delta\bm{D}}\cdot\{\bm{X},\bm{X}\}_s\cdot\frac{\delta\mathcal{B}}{\delta \bm{D}}
    \nonumber
    \\
    &
    +4\pi c\int_{X}\left(\frac{\delta\mathcal{A}}{\delta \bm{D}}\cdot\nabla\times\frac{\delta\mathcal{B}}{\delta\bm{B}_1}-\frac{\delta\mathcal{B}}{\delta \bm{D}}\cdot\nabla\times\frac{\delta\mathcal{A}}{\delta\bm{B}_1}\right).
\end{align}
The notation $\int_Z$ refers to integration over the phase-space coordinates $(\bm{X},p_\parallel,\mu,\theta)$ while $\int_{X}$ and $\int_P$ refer to integration only over the spatial and velocity extent, respectively. Notably, the distributional densities $F_s$ contain the phase-space Jacobians: $\int_P F_s(\bm{Z})$ is the number of gyrocenters within a volume element $d\bm{X}$. The single-particle Poisson bracket $\{\cdot,\cdot\}$ in \eqref{eq:functional_poisson_bracket} can be evaluated with respect to any two phase-space functions according to
\begin{align}\label{eq:particle_poisson_bracket}
    \{f,g\}&=\frac{q}{mc}\left(\frac{\partial f}{\partial \theta}\frac{\partial g}{\partial \mu}-\frac{\partial f}{\partial \mu}\frac{\partial g}{\partial \theta}\right)
    -\frac{c\bhat_0}{qB_\parallel^\ast}\cdot\nabla^\ast f\times\nabla^\ast g
    \nonumber
    \\
    &
    +\frac{\bm{B}^\ast}{B_\parallel^\ast}\cdot\left(\nabla^\ast f\frac{\partial g}{\partial p_\parallel}-\frac{\partial f}{\partial p_\parallel}\nabla^\ast g\right).
\end{align}
The terms $\bm{B}^\ast=\nabla\times\bm{A}^\ast$ and $B_\parallel^\ast=\bm{B}^\ast\cdot\bhat_0$ are constructed from the so-called modified vector potential 
\begin{align}\label{eq:A_star}
    \bm{A}^\ast&=\bm{A}_0+(p_\parallel c/q)\bhat_0-(mc^2/q^2)\mu\bm{R}_0^\ast+\bm{A}_1 \equiv\bm{A}_0^\ast+\bm{A}_1,
\end{align}
in the usual way with $\bhat_0=\bm{B}_0/|\bm{B}_0|$ the background magnetic field unit vector. The modified gradient operator is $\nabla^\ast=\nabla+\bm{R}_0^\ast\;\partial/\partial\theta$,
where $\bm{R}_0^\ast=\bm{R}_0+\frac{1}{2}\nabla\times\bhat_0$, and Littlejohn's gyrogauge vector $\bm{R}_0$ is constructed from the background magnetic field. The label $s$ is needed to remain mindful of the species-dependent particle mass and charge. 

In the absence of collisions, the Hamiltonian functional \eqref{eq:hamiltonian_functional} and the functional Poisson bracket \eqref{eq:functional_poisson_bracket} determine the temporal evolution of any functional $\Psi[F,\bm{D},\bm{B}_1]$ via the differential equation
\begin{align}
    \frac{d\Psi}{dt}=[\Psi,\mathcal{H}].
\end{align}
Because the bracket \eqref{eq:functional_poisson_bracket} is antisymmetric, the Hamiltonian $\mathcal{H}$ is trivially conserved $d\mathcal{H}/dt=[\mathcal{H},\mathcal{H}]=0$. As demonstrated explicitly in \cite{Brizard_bracket_2021}, the toroidal angular momentum functional
\begin{align}\label{eq:toroidal_momentum_functional}
    \mathcal{P}_\varphi=\sum_s\int_ZF_sp_{\varphi 0,s}+\frac{1}{4\pi c}\int_X\bm{D}\times\bm{B}_1\cdot(\hat{\bm{z}}\times \bm{X}),
\end{align}
where $p_{\varphi 0}=(q/c)\bm{A}_0^\ast\cdot(\hat{\bm{z}}\times\bm{X})$ is the guiding-center single-particle angular momentum, is also conserved, i.e.,  $d\mathcal{P}_\varphi/dt=[\mathcal{P}_\varphi,\mathcal{H}]=0$, on the condition that the background magnetic field $\bm{B}_0$ is axially symmetric. Finally, the entropy functional
\begin{align}
    \mathcal{S}[F,\bm{B}_1]=-\sum_s\int_ZF_s\ln\left(F_s/B_{\parallel s}^\ast\right),
\end{align}
is a Casimir invariant of the functional Poisson bracket, i.e., $[\mathcal{S},\mathcal{A}]=0$ with respect to any arbitrary functional $\mathcal{A}$. For details regarding the derivation of these results, see \cite{Brizard_bracket_2021}. 

Previously in \cite{Hirvijoki_Burby_collisions:2020PhPl}, a symmetric, so-called metric bracket representative of collisions was found for electrostatic gyrokinetic theory. The discovery was based on first constructing a weak formulation for the collision operator presented in \cite{Burby_Brizard_Qin_collisions:2015PhPl} and then symmetrizing the result. This guaranteed that both the system's Hamiltonian and toroidal angular momentum functionals were Casimirs of the resulting metric bracket and provided a metriplectic formulation of collisional electrostatic gyrokinetics. The same strategy cannot be directly applied here: no proper collision operator for electromagnetic gyrokinetics has been presented yet. The work \cite{Hirvijoki_Burby_collisions:2020PhPl} nevertheless provides valuable directions. Following them, we seek a symmetric positive semi-definite functional bracket with the structure
\begin{widetext}
\begin{align}\label{eq:metric_bracket}
    (\mathcal{A},\mathcal{B})=\frac{1}{2}\sum_{s\overline{s}}\int_{Z}\int_{\overline{Z}}F_s(\bm{Z})F_{\overline{s}}(\overline{\bm{Z}})\,  \bm{\Gamma}_{s\overline{s}}(\mathcal{A};\bm{Z},\overline{\bm{Z}})\cdot
    {\sf Q}_{s\overline{s}}(\bm{Z},\overline{\bm{Z}})\cdot\bm{\Gamma}_{s\overline{s}}(\mathcal{B};\bm{Z},\overline{\bm{Z}}),
\end{align}
\end{widetext}
where the 3-by-3 matrix ${\sf Q}_{s\overline{s}}(\bm{Z},\overline{\bm{Z}})$ is defined as 
\begin{align}\label{eq:metric_matrix}
    {\sf Q}_{s\overline{s}}(\bm{Z},\overline{\bm{Z}})=\nu_{s\overline{s}}\,\delta_{s\overline{s}}(\bm{Z},\overline{\bm{Z}})\,\mathbb{Q}(\bm{\Gamma}_{s\overline{s}}(\mathcal{H};\bm{Z},\overline{\bm{Z}})).
\end{align}
The localizing delta-function $\delta_{s\overline{s}}(\bm{Z},\overline{\bm{Z}})=\delta^3(\bm{y}_{s}(\bm{Z})-\bm{y}_{\overline{s}}(\overline{\bm{Z}}))$ enforces the collisions to be local in spatial coordinates via the position $\bm{y}_s(\bm{Z})=\bm{X}+\bm{\rho}_{0s}$ of a particle of species $s$ where $\bm{\rho}_{0}$ is the lowest-order Larmor radius evaluated in terms of the background magnetic field. The Landau matrix $\mathbb{Q}(\bm{\xi})=|\bm{\xi}|^{-1}(\mathbb{I}-\bm{\hat{\xi}}\bm{\hat{\xi}})$ is a scaled projection matrix and the coefficient $\nu_{s\overline{s}}=2\pi q_s^2 q_{\overline{s}}^2\ln\Lambda_{s\overline{s}}$ contains the Coulomb logarithm $\Lambda_{s\overline{s}}$, both familiar from the Landau collision operator. The bracket structure \eqref{eq:metric_bracket} and \eqref{eq:metric_matrix}, regardless of the expression for $\bm{\Gamma}_{s\overline{s}}(\mathcal{A};\bm{Z},\overline{\bm{Z}})$, guarantees that the Hamiltonian functional $\mathcal{H}$ is a Casimir invariant of the metric bracket: $(\mathcal{H},\mathcal{A})=0$, with respect to any functional $\mathcal{A}$. This can be credited to the projection property $\bm{\Gamma}_{s\overline{s}}(\mathcal{H};\bm{Z},\overline{\bm{Z}})\cdot\mathbb{Q}(\bm{\Gamma}_{s\overline{s}}(\mathcal{H};\bm{Z},\overline{\bm{Z}}))=\bm{0}$.

The remaining question then concerns the choice of the vector-valued operator $\bm{\Gamma}_{s\overline{s}}(\mathcal{A};\bm{Z},\overline{\bm{Z}})$. In the electrostatic case \cite{Burby_Brizard_Qin_collisions:2015PhPl,Hirvijoki_Burby_collisions:2020PhPl}, the particle velocity needed in the matrix $\mathbb{Q}$ was defined as $\dot{\bm{y}}_s=\{\bm{y}_s,(\delta \mathcal{H}/\delta F_s)\}_s$, with the bracket \eqref{eq:particle_poisson_bracket}, and the conservation of toroidal angular momentum was guaranteed by the identity $\{\bm{y}_s,p_{\varphi 0}\}=\bm{\hat{z}}\times\bm{y}_s$. In electromagnetic theory, the time dependence in the symplectic part of the gyrocenter Lagrangian also contributes to the particle velocity and the identity crucial for momentum conservation in the electrostatic case no longer applies. The issue was further discussed in \cite{Iorio_Hirvijoki:2021JPlPh}, unfortunately to no avail. This puzzle begins to unravel upon using a definition for the particle velocity that is compatible with the Hamiltonian formulation of the gyrokinetic Vlasov--Maxwell system, namely
\begin{align}\label{eq:y_dot}
    \frac{d\bm{y}_s}{dt}=\left\{\bm{y}_s,\frac{\delta \mathcal{H}}{\delta F_s}\right\}_s+4\pi q_s\frac{\delta \mathcal{H}}{\delta\bm{D}}\cdot\{\bm{X},\bm{y}_s\}_s,
\end{align}
and simultaneously discovering the identity
\begin{align}
    \frac{d\bm{y}_s}{d\varphi} = \left\{\bm{y}_s,\frac{\delta \mathcal{P}_\varphi}{\delta F_s}\right\}_s+4\pi q_s\frac{\delta \mathcal{P}_\varphi}{\delta\bm{D}}\cdot\{\bm{X},\bm{y}_s\}_s=\bm{\hat{z}}\times\bm{y}_s.
\end{align}
We note that this identity holds even if the modified gyrogauge vector $\bm{R}_0^\ast$ is dropped from the modified vector potential \eqref{eq:A_star}, as is often customary. In the electrostatic case, the vector needed in the metric bracket was defined as $\bm{\Gamma}_{s\overline{s}}(\mathcal{A};\bm{Z},\overline{\bm{Z}})=\{\bm{y}_s,\delta\mathcal{A}/\delta F_s\}_s(\bm{Z})-\{\bm{y}_{\overline{s}},\delta\mathcal{A}/\delta F_{\overline{s}}\}_{\overline{s}}(\overline{\bm{Z}})$, based on the weak formulation of the collision operator and symmetrization of it. We, therefore, propose the following modified expression
\begin{widetext}
\begin{align}
    \bm{\Gamma}_{s\overline{s}}(\mathcal{A};\bm{Z},\overline{\bm{Z}})&=\left\{\bm{y}_s,\frac{\delta \mathcal{A}}{\delta F_s}\right\}_s(\bm{Z})+4\pi q_s\frac{\delta \mathcal{A}}{\delta\bm{D}(\bm{X})}\cdot\{\bm{X},\bm{y}_s\}_s(\bm{Z})
    -\left\{\bm{y}_{\overline{s}},\frac{\delta \mathcal{A}}{\delta F_{\overline{s}}}\right\}_{\overline{s}}(\overline{\bm{Z}})-4\pi q_{\overline{s}}\frac{\delta \mathcal{A}}{\delta\bm{D}(\overline{\bm{X}})}\cdot\{\bm{X},\bm{y}_{\overline{s}}\}_{\overline{s}}(\overline{\bm{Z}}).
\end{align}
\end{widetext}
Hence, from Eq.~\eqref{eq:y_dot},  $\bm{\Gamma}_{s\overline{s}}(\mathcal{H};\bm{Z},\overline{\bm{Z}})=\dot{\bm{y}}_s(\bm{Z})-\dot{\bm{y}}_{\overline{s}}(\overline{\bm{Z}})$ becomes the desired difference in the particle velocities of species $s$ and $\overline{s}$ required in the matrix $\mathbb{Q}$ of the Landau collision operator. Further, now $\bm{\Gamma}_{s\overline{s}}(\mathcal{P}_\varphi;\bm{Z},\overline{\bm{Z}})=\bm{\hat{z}}\times(\bm{y}_s(\bm{Z})-\bm{y}_{\overline{s}}(\overline{\bm{Z}}))$ which, together with $\delta_{s\overline{s}}(\bm{Z},\overline{\bm{Z}})$ in the matrix \eqref{eq:metric_matrix}, guarantees that the toroidal momentum functional \eqref{eq:toroidal_momentum_functional} is a Casimir invariant of the metric bracket \eqref{eq:metric_bracket} in the sense of $(\mathcal{P}_\varphi,\mathcal{A})=0$, with respect to an arbitrary functional $\mathcal{A}$.

The new metriplectic formulation for the gyrokinetic Vlasov--Maxwell--Landau theory therefore evolves arbitrary functionals $\Psi[F,\bm{D},\bm{B}_1]$ according to the differential equation
\begin{align}\label{eq:metriplectic_dynamics}
    \frac{d\Psi}{dt}=[\Psi,\mathcal{H}]+(\Psi,\mathcal{S}).
\end{align}
This guarantees energy conservation $d\mathcal{H}/dt=[\mathcal{H},\mathcal{H}]+(\mathcal{H},\mathcal{S})=0$ and, in an axially symmetric magnetic background field, toroidal angular momentum conservation $d\mathcal{P}_\varphi/dt=[\mathcal{P}_\varphi,\mathcal{H}]+(\mathcal{P}_\varphi,\mathcal{S})=0$ on the basis of both $\mathcal{H}$ and $\mathcal{P}_\varphi$ being Casimirs of the metric bracket. The formalism also guarantees monotonic entropy production $d\mathcal{S}/dt=[\mathcal{S},\mathcal{H}]+(\mathcal{S},\mathcal{S}) = (\mathcal{S},\mathcal{S}) \geq 0$ on the basis of $\mathcal{S}$ being a Casimir of the Poisson bracket and the metric-bracket being positive semi-definite.

The kinetic equation for the test-particle phase-space density $F_s$ is found by choosing a functional $\Psi(\bm{Z},t)=\int_{Z'}\delta^6(\bm{Z}'-\bm{Z}) F_s(\bm{Z}',t)$ and evaluating the equation $\partial_t\Psi=[\Psi,\mathcal{H}]+(\Psi,\mathcal{S})$. This results in 
\begin{align}\label{eq:kinetic_equation}
    \partial_tF_s+\partial_\alpha(F_s V_s^\alpha)=\sum_sC_{s\overline{s}}[F_s,F_{\overline{s}}],
\end{align}
where $V^\alpha=\{Z^\alpha,K\}+q\bm{E}_1\cdot\{\bm{X},Z^\alpha\}$ is the Hamiltonian phase-space vector field and the non-linear collision operator is given by
\begin{align}\label{eq:collision_operator}
    C_{s\overline{s}}[F_s,F_{\overline{s}}]&=-\partial_\alpha(\bm{\gamma}_{s\overline{s}}\cdot\{\bm{y}_s,Z^\alpha\}_s)
    \nonumber\\&
    =\partial_\alpha\Big(B_{\parallel s}^\ast D_{s\overline{s}}^{\alpha\beta}\partial_\beta \big(F_s/B_{\parallel s}^\ast\big)-K_{s\overline{s}}^\alpha F_s\Big). 
\end{align}
The collisional-flux-related term $\bm{\gamma}_{s\overline{s}}$, a three-component vector
\begin{align}\label{eq:gamma_flux}
     \bm{\gamma}_{s\overline{s}}(\bm{Z})&=\int_{\overline{Z}}{\sf Q}_{s\overline{s}}(\bm{Z},\overline{\bm{Z}})\,F_s(\bm{Z})F_{\overline{s}}(\overline{\bm{Z}})\cdot\bm{\Gamma}_{s\overline{s}}(\mathcal{S};\bm{Z},\overline{\bm{Z}})
    \nonumber\\&
     = F_{s}{\bf K}_{s\overline{s}}-B_{\|s}^{*}\{\bm{y}_{s}, F_{s}/B_{\|s}^{*}\}_{s}
     \cdot {\sf D}_{s\overline{s}},
\end{align}
and the phase-space diffusion and friction coefficients
\begin{align}
    D_{s\overline{s}}^{\alpha\beta}(\bm{Z}) &= \{\bm{y}_{s}, Z^{\alpha}\}_{s}\cdot{\sf D}_{s\overline{s}}(\bm{Z})\cdot
    \{\bm{y}_{s}, Z^{\beta}\}_{s},\\
    K^{\alpha}_{s\overline{s}}(\bm{Z}) &= \{\bm{y}_{s}, Z^{\alpha}\}_{s}\cdot{\bf K}_{s\overline{s}}(\bm{Z}),
\end{align} 
are expressed in terms of the guiding-center and gyrocenter phase-space transformations of the Fokker-Planck diffusion and friction coefficients that are functionals of the field-particle density $F_{\overline{s}}$ (and the electromagnetic fields)
\begin{align}
    \label{eq:D_3x3}
    {\sf D}_{s\overline{s}}(\bm{Z})&=\int_{\overline{Z}}{\sf Q}_{s\overline{s}}(\bm{Z},\overline{\bm{Z}})F_{\overline{s}}(\overline{\bm{Z}}),\\
    \label{eq:K_3}
    {\bf K}_{s\overline{s}}(\bm{Z})&=\int_{\overline{Z}}{\sf Q}_{s\overline{s}}(\bm{Z},\overline{\bm{Z}})\cdot B_{\parallel,\overline{s}}^\ast(\overline{\bm{Z}}) \big\{\bm{y}_{\overline{s}},F_{\overline{s}}/B_{\parallel,\overline{s}}^\ast\big\}_{\overline{s}}(\overline{\bm{Z}}).
\end{align}
In the electrostatic limit, the result agrees with the gyrokinetic collision operator summarized in Eqs. 22--25 in \cite{Burby_Brizard_Qin_collisions:2015PhPl}, evident from the expressions \eqref{eq:gamma_flux}, \eqref{eq:D_3x3}, and \eqref{eq:K_3}. In the absence of fluctuations, expressions for the phase-space diffusion and friction coefficients were given in \cite{Brizard_2004PhPl} for Maxwellian field-particle distributions in a nonuniform background magnetic field: The spatial diffusion coefficient $D^{\bm{X}\bm{X}} =(D_\mu B/m\Omega^{2}+ D_\perp/m^{2}\Omega^{2})(\mathbf{1}-\bhat\bhat)$, where $D_\mu =\mu (D_\parallel-D_\perp)/(2m E)$ and $D_E = D_\parallel/m$, represents classical transport in a magnetized plasma \cite{Helander_Sigmar}, while the spatial components $(K^{\bm {X}},D^{\bm{X}\mu},D^{\bm{X}E}) = (\nu, D_{\mu},D_{E})\,\bhat\times\bm{v}_{gc}/\Omega$ are orientated in the direction of the guiding-center polarization shift $\bhat\times\bm{v}_{gc}/\Omega$ involving magnetic gradient and curvature in non-uniform magnetic field. 

As the metric bracket \eqref{eq:metric_bracket} operates on functionals depending on $\bm{D}$, it also contributes to the gyrokinetc Maxwell-Amp\`ere equation. Choosing a test functional $\Psi(\bm{X},t)=\int_{X'}\delta^3(\bm{X}-\bm{X}')\bm{D}(\bm{X}',t)$, and evaluating $\partial_t\Psi=[\Psi,\mathcal{H}]+(\Psi,\mathcal{S})$, we find
the gyrokinetic Maxwell-Amp\`ere equation
\begin{align}\label{eq:maxwell_ampere}
    \frac{1}{c}\frac{\partial\bm{D}}{\partial t}+\frac{4\pi}{c}\sum_s\int_Pq_sF_sV^{\bm{X}}_s+\frac{4\pi}{c}\bm{j}_C=\nabla\times\bm{H},
\end{align}
where $\bm{H}$ is defined in the standard manner
\begin{align}
    \bm{H}=\bm{B}_0+\bm{B}_1+4\pi\frac{\delta \mathcal{K}}{\delta \bm{B}_1}.
\end{align}
and the collisional contribution to the current density, $\bm{j}_C$, is given by
\begin{align}
    \bm{j}_C&=\sum_{s\overline{s}}\int_Pq_s\bm{\gamma}_{s\overline{s}}\cdot\{\bm{y}_s,\bm{X}\}_s\nonumber\\
    &=\sum_{s\overline{s}}\int_Pq_s\Big(K_{s\overline{s}}^{\bm{X}} F_s-B_{\parallel s}^\ast D_{s\overline{s}}^{\bm{X}\beta}\partial_\beta \big(F_s/B_{\parallel s}^\ast\big)\Big).
\end{align}
Notably, the new collisional term is mandatory to guarantee that the Maxwell--Amp\`ere equation remains consistent with the Gauss's law for the displacement field: taking spatial divergence of the new Maxwell--Amp\`ere equation \eqref{eq:maxwell_ampere}, and using the kinetic equation \eqref{eq:kinetic_equation} and the collision operator \eqref{eq:collision_operator}, results in
\begin{align}\label{eq:gauss_law}
    \nabla\cdot\bm{D}=4\pi\sum_s\int_Pq_sF_s.
\end{align}
Consequently, the gyrokinetic charge conservation law also has the same new collisional contribution. Indeed, by considering the gyrocenter charge-density functional $\varrho(\bm{X},t) = \sum_{s} \int_{Z'} \delta^{3}(\bm{X}-\bm{X}')\,q_s\,F_{s}(\bm{Z}')$, evaluating $\partial_t\varrho=[\varrho,\mathcal{H}]+(\varrho,\mathcal{S})$
results in
\begin{equation}
\frac{\partial\varrho}{\partial t} +\nabla\cdot\bigg( \sum_s\int_Pq_sF_sV^{\bm{X}}_s+ \bm{j}_C\bigg)=0.
\end{equation}
The Maxwell--Faraday equation is derived by choosing a functional $\Psi(\bm{X},t)=\int_{X'}\delta^3(\bm{X}-\bm{X}')\bm{B}_1(\bm{X}',t)$ which provides the standard expression
\begin{align}
    \frac{1}{c}\frac{\partial \bm{B}_1}{\partial t}+\nabla\times\bm{E}_1=0.
\end{align}

The new formulation presented in this letter establishes a theoretical foundation for gyrokinetic Vlasov--Maxwell--Landau theory. Exploiting the metriplectic formalism enabled not only derivation of the gyrokinetic Landau operator but also retaining the energy and momentum conservation accompanied with monotonic entropy production. While it has long been understood that collisions in the gyrocenter coordinates affect also the spatial $\bm{X}$ coordinates \cite{Xu_Rosenbluth:PoF1991,Brizard_2004PhPl}, the new metriplectic formalism uncovers the resulting implications for the Maxwell-Amp\`ere equation and the charge conservation law. It is expected that the new theory could be useful also in finding structure-preserving discretizations, similarly as in \cite{hirvijoki:2021PPCF}.

The work of EH was supported by the Academy of Finland grant no. 315278. The work of JWB was supported by the Los Alamos National Laboratory LDRD program under project number 20180756PRD4. The work by AJB was supported by the National Science Foundation grant no. PHY-1805164.

\bibliography{references}

\end{document}